\documentclass[aps,prl,twocolumn,superscriptaddress,showpacs,10pt]{revtex4-1}

\usepackage{amsmath}
\usepackage{bm}
\usepackage{graphicx}
\usepackage{subfigure}
\usepackage[ansinew]{inputenc}
\usepackage[colorlinks=true,citecolor=blue,linkcolor=red,linktocpage=true,pagebackref=false]{hyperref}

\newcommand{\ud}{\mathrm{d}}
\newcommand{\p}{\partial}

\newcommand{\OR}{\Omega}

\newcommand{\Ht}{\mathcal{H}} 
\newcommand{\hb}{\hat{a}} 
\newcommand{\hbd}{\hat{a}^{\dagger}} 
\newcommand{\hn}{\hat{n}} 

\newcommand{\bt}[1]{\left<#1\right>} 
\newcommand{\ket}[1]{\left|#1\right>} 

\newcommand{\parit}[1]{{\par\it #1.---}}
\begin{document}

\title{An On-Demand Single-Electron Time-Bin Qubit Source}


\author{J.~R.~\surname{Ott}}\email{johan.ott@unige.ch}
\affiliation{Département de Physique Théorique, Université de Genève, CH-1211 Genève 4, Switzerland}

\author{M.~\surname{Moskalets}}
\affiliation{Département de Physique Théorique, Université de Genève, CH-1211 Genève 4, Switzerland}
\affiliation{Department of Metal and Semiconductor Physics, National Technical University ``Kharkiv Polytechnic Institute", 61002 Kharkiv, Ukraine}


\date{\today}

\begin{abstract}
\noindent We propose a source capable of on-demand emission of single electrons with a wave packet of controllable shape and phase. The source consists of a hybrid quantum system, relying on currently experimentally accessible components. We analyze in detail the emission of single electron time-bin qubits, which we characterize using the well known electronic Hong--Ou--Mandel~(HOM) interferometry scheme. Specifically, we show that, by controlling the phase difference of two time-bin qubits, the Pauli peak, the electronic analogue of the well known optical HOM dip, can be continuously removed. The proposed source constitutes a promising approach for scalable solid-state architectures for quantum operations using electrons and possibly for an interface for photon to electron time-bin qubit conversion. 
\end{abstract}


\maketitle

\parit{Introduction}
One main ingredient in quantum information processes is the quantum bit or qubit~\cite{MANielsen_1st_ed2000}. Like its classical counterpart, the qubit consists of two different states, but contrary to the classical bit, the qubit can be in a superposition of the two states. An example of such is the time-bin qubit of the form $\ket{\psi}=\alpha\ket{0}+\beta e^{i\varphi}\ket{1}$, where $\alpha$, $\beta$, and $\varphi$ are real-valued, $\alpha^2+\beta^2=1$, and $\ket{0}$ and $\ket{1}$ describe two time-bins of a propagating state in which quantum information is encoded. In this Letter we propose a method for creating and characterizing an on-demand single-electron time-bin qubit (SETBQ) with tunable phase difference, $\varphi$. With the progress in hybrid circuit quantum electrodynamics~(QED) in mind~\cite{wallraff_nature04a,schoelkopf_nature08a,franceschi_naturenano10a,you_nature11a,delbecq_prl11a,frey_prl12a,frey_prb12a,petersson_nature12a,basset_prb13a,delbecq_ncomm13a,van_loo_science13a,toida_prl13a,viennot_arxiv13a,deng_arxiv13a,liu_arxiv14a}, we take inspiration from a quantum optics scheme for manipulating the shape and, importantly, also phase of a single-photon wave packet envelope~\cite{law_jmo97a,keller_njp04a,vasilev_njp10a} and transfer it to the electronic realm.

In quantum optics, a single-photon envelope can be shaped using a $\Lambda$-type three-level system, such as an atom or a quantum dot (QD), with one transition resonantly driven by a time-dependent driving field, while the other transition is coupled to a cavity~\cite{law_jmo97a,keller_njp04a}. By controlling the temporal dependence of the phase and amplitude of the driving field, a single photon is emitted into the cavity with an envelope having a time-dependent phase and amplitude. This scheme was initially proposed and demonstrated for three-level quantum emitters in free space~\cite{law_jmo97a,kuhn_prl02a,keller_nature04a,keller_njp04a,vasilev_njp10a,nisbet_jones_njp11a}, for which the creation and characterization of high quality single-photon time-bin qubits has been achieved~\cite{vasilev_njp10a,nisbet_jones_njp13a}. Recently, the shaping of single-photon envelopes has also been demonstrated with solid-state QDs at optical frequencies~\cite{matthisen_naturecom13a} and at microwave frequencies~\cite{pechal_arxiv13a,wenner_arxiv13a}.

\begin{figure}[ht!]
	\centering
		\includegraphics[width=0.45\textwidth]{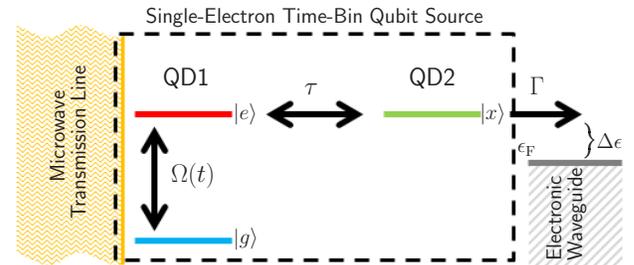}
	\caption{Color online.  A schematic representation of the setup. A time-dependent driving field, $\OR(t)$, in the microwave transmission line (left orange region) is coupled to QD1, a two-level quantum dot with ground level $\ket{g}$ (blue line) and excited level $\ket{e}$ (red line). QD1 is in turn tunnel coupled, with amplitude $\tau$, to QD2, a quantum dot with one level $\ket{x}$ (green line) from which the electron may escape, with rate $\Gamma$, into the electronic waveguide (right gray region). The dashed rectangle shows the components of the single-electron time-bin qubit source.}
	\label{fig:fig1_setup}
\end{figure}

Within quantum electronics, several types of single-electron sources (SES) are realized experimentally, for a review see Ref.~\cite{pekola_rmp2013a}, while electron emission using an optically driven double quantum dot has been proposed theoretically~\cite{stafford_prl96a}, for a review see Ref.~\cite{platero_pr04a}, and recently also using circuit QED~\cite{van_den_berg_arxiv14a}. However, for an electronic flying qubit, controlling the electron phase is crucial. The earliest efforts to electrically control an electron phase in solid-state circuits required strong magnetic fields~\cite{washburn_prl87a,de_vegvar_prb89a,van_oudenaarden_nature98a}. The most recent demonstration combines an Aharonov-Bohm ring with a two-channel wire to generate solid-state flying qubits~\cite{yamamoto_nnano12a}. The later setup is limited by backscattering at the ring-wire interface, resulting in a visibility of less than $1\%$. Here we propose a different approach, which allows to efficiently produce a highly controllable time-bin single-electron qubit. We suggest to employ a hybrid quantum system, see Fig.~\ref{fig:fig1_setup}, where a microwave driving field is used to raise an electron from the ground state of a double quantum dot to the excited state above the Fermi level of a nearby electronic waveguide. The essential ingredient of our proposal is a controlled temporal dependence of the amplitude and phase of the driving field, that allows to produce the time-dependent amplitude and phase of an emitted electron. The experimental toolbox for such time-dependent manipulation of the microwave driving field was recently demonstrated for the amplitude~\cite{pechal_arxiv13a,wenner_arxiv13a} and phase~\cite{hoi_prl13a,abdo_prl13a,tancredi_apl13a}.

To demonstrate coherence properties of single particles one may use one of several characterization schemes, \textit{e.g.} Hanbury Brown--Twiss~\cite{hanbury_brown_nature56a,hanbury_brown_nature56b}~(HBT), Mach--Zehnder~\cite{zehnder_zfi81a,mach_zfi82a}~(MZ), and Hong--Ou--Mandel~\cite{hong_prl87a}~(HOM) interferometry. These schemes were originally developed for optics, but have within the last two decades also been realized in electronics~\cite{henny_science99a,oliver_science99a,ji_nature03a,samuelsson_prl04a,neder_prl06a,neder_nature07a,samuelsson_prl09a,bocquillon_prl12a,dubois_nature_13a,bocquillon_science13a,dubois_nature_13a}. The coherence properties of electrons emitted on-demand~\cite{feve_science07a,dubois_nature_13a} were recently characterized via electronic Hong-Ou-Mandel interferometry~\cite{bocquillon_science13a,dubois_nature_13a}. These experiments demonstrate the possibility to achieve correlations between electrons emitted by independent sources and, for instance, open the opportunity to generate time-bin entangled electron pairs as suggested in Ref.~\cite{splettstoesser_prl09a}. Other recent proposals for such states utilize the helical edge states of a quantum spin Hall insulator~\cite{inhofer_prb13a,hofer_prb13a}.

In the following we first describe the setup and model of a source capable of creating SETBQs and then analyze its characterization using HOM interferometry.

\parit{Setup} 
We consider a two-level quantum dot, QD1, tunnel coupled to a single-level quantum dot, QD2, which is coupled to an electronic waveguide, \textit{i.e.} a ballistic conductor or an edge state, see Fig.~\ref{fig:fig1_setup}. A microwave transmission line in the vicinity of QD1 allows the ground state, $\ket{g}$, and exited state, $\ket{e}$, to be coupled via a classical time-dependent microwave field, $\OR(t)$. An electron in $\ket{e}$ may tunnel with amplitude $\tau$ into the state $\ket{x}$ of QD2 from which it can escape with rate $\Gamma$ into the electron waveguide at an energy state $\Delta \epsilon$ above the Fermi energy, $\epsilon_F$. The state $\ket{x}$ of QD2 ensures that the electrons below the Fermi level of the electronic waveguide do not couple to QD1. While the analysis does not refer to a specific experimental setup, the scheme could be realized in a variety of systems such as by discrete levels of a carbon nanotube, where coupling to fermionic leads and a microwave circuit cavity has been realized~\cite{delbecq_prl11a,delbecq_ncomm13a} or gate defined QDs coupled to microwave transmission lines as \textit{e.g.} investigated in Refs.~\cite{frey_prl12a,frey_prb12a,basset_prb13a}.

\parit{Model} 
The Hamiltonian describing a quantum dot interacting with a microwave transmission line is well known~\cite{childress_pra04a,blais_pra04a,bergenfeldt_prb12a}. We assume an infinite on-site Coulomb interaction, such that the source is at most occupied by one electron, consider low temperature, such that no electrons leak from the electronic waveguide into the source, and treat the escape rate by a dissipative Lindblad term of state $\ket{x}$. Thereby, in the interaction picture, the coherent evolution of a single electron in the source is described by the effective non-Hermitian Hamiltonian, ($\hbar=1$)
\begin{align}
\Ht=(\OR(t)\hbd_e\hb_g+\tau\hbd_e\hb_x+H.c.)-i\Gamma\hbd_x\hb_x,\label{eqn:H'}
\end{align}
where the operators $\hb_i$ and $\hbd_i$ annihilate and create an electron in state $i$. For the time being we analyze the idealized situation and disregard decoherence effects from relaxation and dephasing for clarity, while we return to these important effects later.

In the basis of $\ket{g}$, $\ket{e}$, and $\ket{x}$, this gives the Schrödinger equation
\begin{align}
\frac{\p}{\p t}
\left[
\begin{array}{c}
a_{g}\\
a_{e}\\
a_{x}
\end{array}
\right]
=
\left[
\begin{array}{ccc}
0        & -i\OR^*(t) & 0 \\
-i\OR(t) & 0          & -i\tau \\
0        & -i\tau^*      & -\Gamma
\end{array}
\right]
\left[
\begin{array}{c}
a_{g}\\
a_{e}\\
a_{x}
\end{array}
\right].\label{eqn:schr}
\end{align}
The amplitudes $a_i$ of states $i=g, e, x$ are related to the density matrix elements $\rho_{ij}$ through $\rho_{ij}=a_ia_j^*$.

Equation~\eqref{eqn:H'} is identical to the effective Hamiltonian of the quantum optical analog for shaping a single photon envelope~\cite{law_jmo97a,vasilev_njp10a} described in the introduction. Specifically, our single-electron source is related to the driven $\Lambda$-type 3-level atom single-photon source, by mapping the microwave drive, $\OR(t)$, the tunneling amplitude, $\tau$, and the escape rate, $\Gamma$, respectively to the optical drive, the atom-cavity coupling, and the cavity leakage. We may thus draw a parallel to the approach for photon shaping to design the single-electron wave packet.

\parit{Creating Single-Electron Time-Bin Qubits}
The electron that coherently escapes the source, propagates in the $z$-direction along the electronic waveguide with a single electron wave packet of the form $\Psi(t,z)=\psi(t,z)e^{i(\epsilon_F+\Delta\epsilon)t-ik z}$. Here, the envelope $\psi(t,z)$ is determined by the rate $\Gamma$ at which the population of the state $\ket{x}$ decays, \textit{i.e.} at $z_S$, the position of the source, $\psi$ is related to the density matrix element $\rho_{xx}(t)$ through $|\psi(t,z_S)|^2= 2\Gamma\rho_{xx}(t)$, and thereby, $a_{x}(t)=\psi(t,z_S)/\sqrt{2\Gamma}$. In the following we omit the explicit dependence of $z$ for simplicity. Similar to the optical analog~\cite{vasilev_njp10a}, by solving Eq.~\eqref{eqn:schr} we can derive an equation, which determines the driving field, $\OR(t)$, to be imposed in order to emit a specified single-electron envelope into an electronic waveguide. Furthermore, for time-bin qubits we find that the time-dependent phase of $\Omega(t)$ directly transfers to the phase of the envelope. 

From the derivation of $\OR(t)$, the shape of the envelope is limited by 
\begin{align}
|\psi(t)|^2\leq2\Gamma, \quad \mbox{and} \quad|\p_t\psi(t)+\Gamma\psi(t)|^2\leq2\Gamma|\tau|^2.\label{eqn:limits}
\end{align} 
These inequalities physically signify that the temporal variation of the envelope cannot be faster than the internal timescales of the source. Specifically, the first inequality results in a limit to the width, $T$, of the envelope as a result of the finite escape rate, $\Gamma$. The second inequality shows that the temporal variation of the envelope is limited by the tunneling rate between the two dots in combination with the escape rate. Moreover, as it is known from the optical analog~\cite{vasilev_njp10a}, the coupling between the driving field and a QD depends on the occupancy of the QD. Thus, since the occupancy of QD1 decreases during the emission process, the amplitude of the driving field has to be increasingly strong to fully emit the single electron. Therefore, in practice, the electron is partially emitted. To account for this effect we introduce the efficiency of the source, $\eta$, and represent $\psi = \sqrt{\eta}\psi_0$, with the efficiency $0< \eta < 1$ and the ideal envelope $\psi_0$ having unit time integral.  These features are best illustrated by an example.

\begin{figure}[t!]
	\centering
		\includegraphics[width=0.45\textwidth]{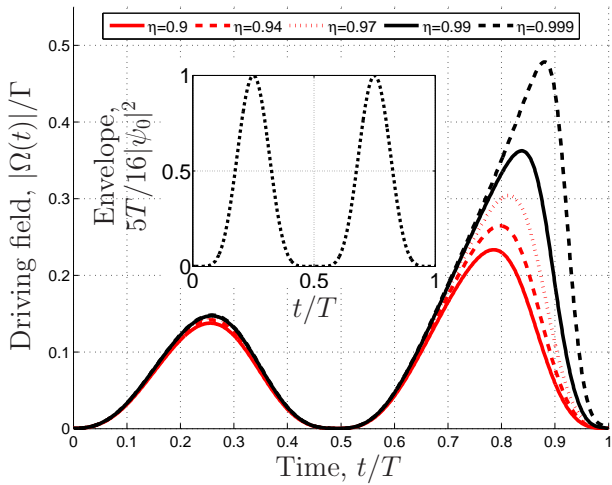}
	\caption{Color online. The calculated temporal shape of the drive $|\OR(t)|/\Gamma$ in order to obtain the single-electron TBQ, Eq.~\eqref{eqn:TBQ}, for different $\eta$. The parameters used for the source are $\tau=2\Gamma$ and $T=100/\Gamma$. Inset: The squared envelope of an emitted electron normalized in maximum to unity.}
	\label{fig:fig2_drive}
\end{figure}

We consider the emission of a time-binned single electron with an envelope $\psi_0(t)=\psi_{sp}(t,0)+\psi_{sp}(t,T/2)e^{i\varphi}$ where $\psi_{sp}(t,t')$ is a semi-pulse given
\begin{align}
\psi_{sp}(t,t')=\frac{4}{\sqrt{5T}}\sin^3\left(\frac{2\pi t}{T}\right), \qquad t'\leq t \leq \frac{T}{2}+t'. \label{eqn:TBQ}
\end{align}
This gives a time-bin qubit of the form $\ket{\psi}=\frac{1}{\sqrt{2}}(\ket{10}+e^{i\varphi}\ket{01})$, where \textit{e.g.} $\ket{10}$ represents an electron in the first semi-pulse. The envelope is shown in the inset of Fig.~\ref{fig:fig2_drive}. For $\tau=2\Gamma$ and $T=100/G$ the temporal shape of the needed field amplitude, $|\OR(t)|$, is shown in Fig.~\ref{fig:fig2_drive} for different $\eta$. The temporal shape of the first part of $\OR(t)$ only changes slightly when increasing $\eta$, while the second part on the other hand has to be increasingly skewed due to the reduced occupancy of QD1. 

We have thus presented a source capable of emitting time-bin qubits. We notice that, even though we have focused on such single electron envelopes, the scheme is not limited to these. In fact any single electron envelope shape is allowed, only restricted by the internal timescales of the source through inequalities Eq.~\eqref{eqn:limits}.

\begin{figure}[t!]
	\centering
		\includegraphics[width=0.45\textwidth]{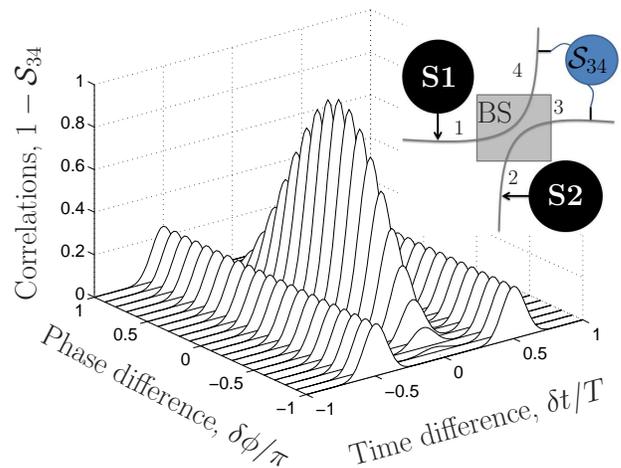}
	\caption{The HOM correlations vs phase difference, $\delta\varphi$, and time difference, $\delta t$. Parameters used are $\tau=2\Gamma$, a desired envelope of width $T=100/\Gamma$, and source efficiency $\eta=0.99$. For $\Omega(t)$, see the solid black line in Fig.~\ref{fig:fig2_drive}. Inset: Sketch of the HOM setup. The two sources S1 and S2 emit single electrons, which collide at the beam splitter BS and the current correlations, $\mathcal{S}_{34}$, between the arms 3 and 4 are measured.\label{fig:fig3_characterization_phi}}
\end{figure}

\parit{Characterization}
Having described a source for creating SETBQs, we next analyze a possible way of characterizing it, specifically using the HOM scheme. The HOM scheme, shown in the inset of Fig.~\ref{fig:fig3_characterization_phi}, consists of a beam-splitter~(BS) having two input ports, 1 and 2, and two output ports, 3 and 4. Electrons are emitted into the input ports followed by the zero-frequency current correlations measurement of the output ports, \textit{i.e.}~\cite{blanter_pr00a} 
\begin{align}
S_{34}&=\iint\ud t\ud t' [\bt{\hn_3(t)\hn_4(t')}-\bt{\hn_3(t)}\bt{\hn_4(t')}],
\end{align}
where $\hn_{\alpha}(t)$ is the current flux operator of channel $\alpha$. If only one of the sources, S1 or S2, is active we get the HBT correlations, $S^{HBT}_{S1}$ or $S^{HBT}_{S2}$, which are identical for identical sources, $S^{HBT}_{S1}=S^{HBT}_{S2}=S^{HBT}$. If both sources are active, the HOM correlations, $S^{HOM}$, are measured and one may define the HOM correlations normalized with respect to the HBT correlations, $\mathcal{S}_{34}=S^{HOM}_{34}/(2S^{HBT})$, to get a quantity which is independent of the transmission and reflection probabilities of the BS and, furthermore, to reduce the effect of temperature~\cite{bocquillon_science13a}. Assuming that none of the energy components of the envelopes overlap with the Fermi sea, we have that~\cite{jonckheere_prb12a}
\begin{align}
\mathcal{S}_{34}&=1-\left|\int\ud t \;\psi_{1}^*(t)\psi_{2}(t)\right|^2. \label{eqn:S_{34}}
\end{align}
In Fig.~\ref{fig:fig3_characterization_phi} we show $\mathcal{S}_{34}$ calculated for two sources, S1 and S2, emitting single electrons with envelopes $\psi_1$ and $\psi_2$ of the form in Eq.~\eqref{eqn:TBQ} and partial phases $\varphi_1$ and $\varphi_2$, respectively, with $\delta\varphi=\varphi_2-\varphi_1$. The sources S1 and S2 are synchronized such that emitted electrons arrive at the beam splitter with a time delay $\delta t$. 

First, for $\delta t=0$, we have $\mathcal{S}_{34}=1-\eta^2\cos^2(\delta\varphi/2)$ and thus for $\delta\varphi=0$, corresponding to two identical incident single-electron envelopes, the current correlation is $\mathcal{S}_{34}=1-\eta^2$. For unit efficiency, $\eta=1$, each electron incoming from the port $1$ encounters an electron incoming from the port $2$. At $\delta t =0$, the perfect overlap of the envelopes results in $\mathcal{S}_{34} = 0$, \textit{i.e.} the electrons are scattered into different output ports $3$ and $4$. This anti-bunching is known as the Pauli peak and reflects the fermionic nature such that two electrons cannot occupy the same state at the same time~\cite{blanter_pr00a,olkhovskaya_prl2008a}. The Pauli peak can be seen in Fig.~\ref{fig:fig3_characterization_phi} at $\delta\varphi =0$ close to $\delta t=0$.

As $\delta\varphi$ changes from zero to $\pi$ we observe that, interestingly, the Pauli peak at $\delta t=0$ disappears. That is, even though the two electrons arrive at exactly the same time, such that $|\psi_1(t)|=|\psi_2(t)|$, they seemingly do not obey the Pauli principle. This peculiar observation is caused by the two envelopes at $\delta\varphi=\pi$ being orthogonal to each other and thus do not constitute the same states, \textit{i.e.} their overlap is zero and electrons can be scattered to the same output port. At equal arrival time, the difference between $\mathcal{S}_{34}$ for $\delta\varphi=0$ and $\delta\varphi=\pi$ shows the difference between the two states $\ket{\psi}=\frac{1}{\sqrt{2}}(\ket{10}\pm\ket{01})$ and thus $\mathcal{S}_{34}$ constitutes the source visibility. 

Lastly, by changing the time-delay, $\delta t$, two phase-independent smaller peaks are seen at $\delta t=\pm T/2$. These correspond to the overlap of the front pulse of $\psi_1(t)$ with the tail pulse of $\psi_2(t)$ and vice versa. We have thus shown that, by varying the time-delay, $\delta t$, and phase difference, $\delta\varphi$, between the two SETBQ, we are able to characterize our proposed source using the HOM scheme.

\parit{Decoherence Effects}
Until now we have neglected the important issues of the relaxation of QD1 and dephasing due to the tunneling between QD1 and QD2. Since we are interested in the coherent evolution of the electron, we follow Ref.~\cite{vasilev_njp10a} and describe the decoherence by a dissipation of the coherent electron to derive an effective Hamiltonian as used in the quantum-jump approach to dissipative systems known from quantum optics, see \textit{e.g.} Refs.~\cite{zoller_pra87a,HCarmichael_1st_1993,plenio_rmp98a}. This is done by including a relaxation rate $\gamma_{r}$ of state $\ket{e}$ and tunneling dephasing rate $\gamma_{\phi}$ of states $\ket{e}$ and $\ket{x}$ as dissipations in the effective Hamiltonian, Eq.~\eqref{eqn:H'}, governing the coherent electron. This implicitly assumes that a photon emitted into the microwave transmission line due to the relaxation of QD1, does not reexcite the QD. We may then again derive an equation, which determines $\OR(t)$ for the coherent emission of a specified single electron envelope into the electronic waveguide. From the conservation of charge one finds that the decoherence terms lead to the physical limit on the maximal efficiency, similar to the optical analog~\cite{vasilev_njp10a},
\begin{align}
\eta_{\max} = \frac{1}{1+\frac{(\gamma_{r}+\gamma_{\phi})}{\Gamma|\tau|^2}\left[\Gamma^2+\int_0^{T}\ud t'|\p_{t'}\psi_0(t')|^2\right]+\frac{\gamma_{\phi}}{\Gamma}},
\end{align}
\textit{i.e.} the efficiency has to be in the interval $0\leq\eta\leq\eta_{\max}$. The case $\eta=\eta_{\max}<1$ signifies that the decoherence results in the emitted electron being in a statistical mixture of being coherently emitted with probability $\eta$ and incoherently emitted with probability $1-\eta$.  To give an estimate of $\eta_{\max}$ we take the currently achievable experimental values $\tau/(2\pi)\sim15.5$GHz, $\gamma_{r}/(2\pi)\sim100$MHz, and $\gamma_{\phi}/(2\pi)\sim 1.5$GHz from Ref.~\cite{basset_prb13a} and $\Gamma \sim \tau/2$ from~\cite{frey_prb12a} and $T=100/G$ as used in Figs.~\ref{fig:fig2_drive} and~\ref{fig:fig3_characterization_phi} giving $\eta_{\max}\sim0.8$ for the SETBQ, Eq.~\eqref{eqn:TBQ}. This suggests that, with current technology, our proposed source has the possibility of a significantly increased efficiency compared to state-of-the-art flying qubit sources~\cite{yamamoto_nnano12a}.

\parit{Reloading the Source}
Lastly, let us describe three possible methods for reloading the source for gate defined QDs. First, a laser pulse could excite electrons from the buffer layer to the state $\ket{g}$ as experimentally demonstrated in Ref.~\cite{fujita_prl13a} and thereby deterministically load the source with a new electron on demand. A second method is to tune the voltage gates to lower the energy level of state $\ket{x}$ below the Fermi sea of the electronic waveguide to the level of $\ket{g}$. This would permit an electron (only one due to the Coulomb energy) to stochastically flow from the electronic waveguide into $\ket{x}$ and then tunnel back and forth between $\ket{g}$ and $\ket{x}$. Since it is below the Fermi sea it would not decay back into the electronic waveguide. Then if $\ket{x}$ is slowly raised, the electron would end in $\ket{g}$. A third option is to couple QD1 very weakly to an electron reservoir. If the time scale of the electron tunneling from the reservoir to QD1 is much longer than the electron emission time of the source, then there would only be a minimal risk of having more than one electron in the emitted electron time-bin.

\parit{Summary and Outlook}
We proposed a single-electron source relying on a hybrid quantum system. With our scheme one can design an electron envelope thereby providing control over the phase and the amplitude of a single electron. In particular, we analyzed the emission of a single-electron time-bit qubit, and showed how to characterize it using a well-known interferometric technique. Specifically, we showed that in Hong--Ou--Mandel interferometry the Pauli peak can be continuously removed by controlling the phase difference of two time-bin qubits. Our analysis showed that, with experimentally relevant parameters, the source efficiency is expected to be close to unity.  This opens the possibility of using single electrons for quantum operations in scalable solid-state architectures. Furthermore, with the recent demonstration of creating single microwave photons with controlled envelopes~\cite{pechal_arxiv13a,wenner_arxiv13a} our proposed scheme constitutes a possible photon-electron interface for photon-to-electron time-bin qubit conversion with microwave coupling in the vacuum Rabi regime.

\parit{Acknowledgment}
We are most grateful to Markus Büttiker who raised the problem addressed here and participated in the initial stage of this work. Markus sadly passed away before the project was finalized. We thank D. Dasenbrook, P. P. Hofer, C. Flindt, P. Samuelsson, and M. Wubs for comments to the manuscript. JRO acknowledges financial support from the Danish Council of Independent Research. Research in Geneva is supported by the Swiss NSF.

%


\end{document}